\documentstyle[emulateapj,astron_mlb]{article}
\begin{document}
\newcommand{\ow}{$W_{\circ}$(OII)}
\newcommand{\mstar}{$M_{K_s}^\ast$}
\newcommand{\jmstar}{$M_{J}^\ast$}
\newcommand{\kalpha}{$\alpha_{K_s}$}
\newcommand{\jalpha}{$\alpha_{J}$}
\def\etal{{\it et al.\thinspace}}
\def\items{\hangindent=0.5truecm \hangafter=1 \noindent}
\newcommand{\shortcite}{\cite*}
\title{The Environmental Dependence of the Infrared Luminosity and Stellar 
Mass Functions$^{1}$}
\author{Michael L. Balogh\altaffilmark{2}}
\affil{Department of Physics, University of Durham\\
 South Road, Durham, UK DH1 3LE}
\author{Daniel Christlein, Ann I. Zabludoff \& Dennis Zaritsky}
\affil{Steward Observatory, University of Arizona\\ 
Tucson, AZ, 85721 USA}
\altaffiltext{1}{This publication makes use of 
data products form the Two Micron All Sky Survey (2MASS), which is a joint project of the University of Massachusetts and 
the Infrared Processing and Analysis Center/California Institute of Technology, funded by the National Aeronautics and
Space Administration and the National Science Foundation.}
\altaffiltext{2}{email:M.L.Balogh@durham.ac.uk}
\begin{abstract}
We investigate the dependence of the galaxy infrared luminosity 
function (LF) and the associated stellar mass function (SMF)
on environment and spectral type using photometry from the
Two Micron All Sky Survey and 
redshifts from the Las Campanas Redshift Survey
for galaxies brighter than
$M_J<-19+5 \mbox{log} h$.  In the field environment, 
galaxies with emission lines
have LFs with much steeper faint end slopes
($\alpha_J=-1.39$) than galaxies without emission lines
($\alpha_J=-0.59$). In the cluster environment, 
however, even the non-emission line
galaxies have a steep faint-end LF ($\alpha_J=-1.22$). There is
also a significant (95\%) difference between the overall cluster
and field LFs, $\Delta \alpha_J=-0.34, \Delta M_J^\ast=-0.54$.
All of these variations are more pronounced in the SMFs, which we compute
by relating the strength of the 4000\AA\ break in the optical spectra to
a stellar mass-to-light ratio.
\end{abstract}
%

\section{Introduction}\large
The shape of the stellar mass function (SMF) is a fundamental property
of the galaxy population because it constrains the baryon content of
galaxies, the fraction of all baryons in stars, and the chemical
evolution of galaxies.  Until recently, however, observational 
work has been unable to provide a reliable, direct estimate of the
SMF.  This failure arises primarily because the conversion of the luminosity function (LF)
into the SMF requires
the poorly-constrained optical mass-to-light (M/L) ratio, which in turn
depends sensitively on both the stellar populations and the 
spatial distribution of dust and stars,
neither of which is known for a substantial sample of
galaxies.  Complex models of dust and star formation are
therefore required to compare the observations with theoretical
predictions (\cite{Cole2000,Granato00}).

It is now possible to
mitigate these difficulties by combining spectroscopic catalogs
with large infrared (IR) surveys.  
The sensitivity of M/L to stellar populations and
extinction, which has plagued the transformation of the optical LF
to a SMF, is lessened at IR wavelengths. M/L varies by only a
factor of $\sim$ two over a large range of star formation histories
(e.g., Gavazzi, Pierini \& Boselli \shortcite{GPB}; Bell \& de Jong
\shortcite{BdJ_ML}; \cite{Cole2000}) and is nearly insensitive to the
presence of dust at the levels expected in most galaxies (the
typical internal extinction at $K_s$ is $\sim0.05$ mag; \cite{Silva}). 

Three recent developments motivate us to attempt the measurement of
the SMF's environmental dependence.
First, large IR surveys are an entirely new
resource, due to the advent of large format detectors, dedicated
survey telescopes, and significant human and material resources.  
Second, the availability and uniform analysis of large
optical and IR surveys has led to a consensus on the shape of the {\sl
global} galaxy LF at both optical and IR wavelengths.  Although
controversy persisted for many years regarding differences between the
optical LFs obtained by different investigators
(e.g., \cite{SAPMI,EEP,marzke_cfa,L+96,Zucca,Ratcliffe}), these
differences have now largely been resolved (Blanton \etal \shortcite{Sloan_lf}).  
At IR wavelengths, agreement among
different samples has been extraordinarily good
(\cite{MSE,GPMC,Gardner,SSCM,Loveday-klf,Kochanek-KLF}; Cole \etal
\shortcite{Cole-2mass}).  This long-awaited concordance makes it
possible to confidently proceed and investigate variations in the LF
for different galaxy types, different environments, and different
redshifts. Third, 
the availability of large, cosmological simulations allow the 
dependence of the SMF on environment to be modeled with some
reliability.  However, the translation from the
dark matter halo mass function to the SMF still depends on 
such complicated and
ill-understood factors as the bias function (\cite{WDEF,NBW}), the gas
cooling rate (\cite{RO}), and reheating due to feedback from star
formation (e.g., Larson \shortcite{Larson74}; Cole \shortcite{Cole91};
Balogh \etal \shortcite{baryons}). 

A dependence of the SMF on galaxy type would not be surprising,
due to the distinctly different star formation histories of late- and
early-type galaxies.  In addition, we can
expect an environmental dependence of the SMF 
because (1) the merging history
(e.g., \cite{LC93}) and bias level (\cite{BBKS,NBW}) of halos in
clusters are different from those of isolated halos, (2) there
are physical processes, such as galaxy harassment
(\cite{harass}), that may operate in dense cluster environments
and alter the halo mass function, and (3) the stellar populations of cluster
galaxies are different from those of field galaxies, perhaps
indicative of different star formation histories
(e.g., \cite{DTS,B+97,B+98,P+99,MW00}).  

Such expectations have already received some support from
observations.  
It is well known that the optical LF depends
on galaxy morphology
(e.g., \cite{SAPMI,BST,marzke_cfa_morph,marzke_morph}), primarily in
the sense that it steepens toward later galaxy types.
Analogous trends are also seen if the population is divided by
emission line strength (\cite{autofib_lf,L+96,SAPMII,Christ}), color
(\cite{marzke_colour,L+96,Treyer}), or principal spectral components
(\cite{Bromley,2dF_lf}).  In addition to the dependence on type,
optical observations also indicate that the dwarf-to-giant ratio
steadily increases with increasing environmental density
(\cite{SBT,Roberto95,FS,Bromley,ZM00,Christ}).  

The trends seen in the optical LFs must be explored in the IR to
determine whether they reflect underlying variations in the SMF or in
M/L.  In one recent example of such work, Kochanek \etal
(\shortcite{Kochanek-KLF}) find that the $K_s-$band LF of
galaxies has the same dependence on morphology as seen at 
optical wavelengths.  This finding is strong evidence that the galactic SMF
is not universal, and that early-type galaxies cannot simply be an
unbiased subsample of late-type galaxies in which star formation has
ceased.  Our primary purpose is to further explore in the IR the trends found in
the optical.
The compilation of the data we use is described in
\S\ref{sec-data}.  In \S\ref{sec-lfunc}, we calculate the $J$ and
$K_s$ band galaxy luminosity functions as a function of environment
and spectral type.  We find that the trends in optical data are also
present in the IR data, and more significantly in the 
stellar mass function.  The implications are explored in
\S\ref{sec-discussion}, and we summarize our conclusions in
\S\ref{sec-conc}.
  
\section{Data}\label{sec-data}
To construct infrared luminosity functions, we use photometry from the
Two Micron All Sky Survey (2MASS; \cite{2MASS}), and redshifts from the 
Las Campanas Redshift Survey (LCRS; \cite{LCRS}).  The
LCRS consists of over 25,000 galaxies with redshifts,
approximately covering the magnitude range $R_c=15$ to $R_c=17.7$,
where $R_c$ is the ``hybrid'' Kron-Cousins $R$ system described in Lin
\etal\ (\shortcite{L+96}).  The survey includes some fields observed
with a 50 fiber configuration, and other, later fields observed with a 112
fiber configuration.  The 50 fiber data are problematic because
the sampling fraction is lower and the apparent
magnitude range of the targeted galaxies is typically less than 1.5
magnitudes.  Galaxies with high central surface
brightnesses are excluded from the redshift survey, and this limit is
about 0.7 mag fainter in the 50 fiber data.  To ensure a homogeneous
and maximally complete sample, we restrict ourselves to the 112 fiber
data, as did Lin \etal (\shortcite{L+96}) for their fiducial
computation of the optical LF.

We use the second incremental data release of the 2MASS catalog, which
contains photometry in $J$, $H$ and $K_s$ bands of both extended and
point sources.  The overlap of this release and the 112-fiber LCRS
fields is patchy, covering only about 40\% of the LCRS.
Cross-correlation between the catalogs is done by matching galaxy
astrometric positions.  Following the example of Cole \etal\
(\shortcite{Cole-2mass}), we allow matches between galaxies whose
positions differ by less than 0.75 times the $J$-band Kron radius, to
allow for the greater positional uncertainty of larger galaxies.

\subsection{Spectral Classification} Zabludoff \etal
(\shortcite{Z+96}) measured the rest frame equivalent width of the
[OII]$\lambda$3727\AA\
emission line, \ow, and the strength of the Balmer break, $D_{4000}$.  We use \ow\ to divide
the sample into broad spectral classes: emission line (EL) 
galaxies with \ow$\ge 5$
\AA\ and non-emission line (NEL) galaxies with \ow$<5$\AA.   
Two caveats accompany this division.  First,
the [OII] line correlates with $H\alpha$ emission, albeit with considerable scatter,
suggesting that it is a good indicator of star formation in an average sense (e.g., \cite{K92}).
However, \ow\ also depends on metallicity and ionization, and 
on extinction if dust is non-uniformly distributed (i.e., more prevalent in
HII regions; \cite{CL}).  All of these properties are expected to depend 
on galaxy luminosity. The correlation between
[OII] and H$\alpha$ does indeed have 
a luminosity dependence (\cite{T+99,Jansen01}).
Second, although, the fibers used in the LCRS are
quite large (3\farcs5), Kochanek (\shortcite{Kochanek-LCRS}) suggests that aperture
effects may still be important.  We address this issue in \S\ref{sec-sptype}.

\subsection{Environment Classification}
Group and cluster catalogs have been generated by
Christlein (\shortcite{Christ}) from the LCRS 
using a friends-of-friends algorithm to 
search for overdensities in redshift space.  For each association, the one
dimensional velocity dispersion, $\sigma_1$, is computed.  We define galaxy
clusters to be those with $\sigma_1>400$ km/s, corresponding to a virial
temperature of about 1 keV.  Associations with smaller $\sigma_1$ are
termed ``groups'' and any galaxy not found to be in an overdense region
is assigned to the ``field'' sample.  Only associations with at least three members
are retained in the final catalog; we discuss the possibility of biases related
to this effect both below and in \S\ref{sec-bias}.  

Velocity dispersions computed from small samples are biased
low due to insufficient sampling of the wings of the distribution 
and the poor, biased determination of the mean (\cite{ZM98}).  
This bias results in some fraction of the associations that properly belong
in the ``cluster" sample being incorrectly placed into the ``group'' sample.
Furthermore, identification of groups with low 
velocity dispersions is sensitive to the linking length used in the cluster-finding
algorithm, and such groups
are more likely to be contaminated by interlopers.  This contamination
results in some
groups that are false detections, and the galaxies in those ``groups" should
be properly identified as isolated galaxies.
To the degree that the environments are misclassified, any underlying 
differences among environments will be diluted.

\subsection{Completeness}
We compute the luminosity functions in both the $K_s$ band, which
is commonly used, and the $J$ band, where the 2MASS photometry is deepest.  
In $J$, we use the Kron aperture magnitudes as the best
estimate of the total magnitude.  Cole \etal\ (\shortcite{Cole-2mass})
demonstrate that the most robust $K_s$ magnitudes are obtained by
combining the Kron $J$-magnitudes with the default $J-K_s$ colors.
This nonintuitive result arises because the $J$ images have a higher
signal-to-noise ratio and allow a more accurate determination of the
Kron aperture size.  We adopt this magnitude definition of $K_s$.  A
small fraction (2.2\%) of the galaxies in the LCRS within our
magnitude range (defined below) are misclassified as stars in 2MASS.
For these galaxies, we use the 8\arcsec\ aperture magnitudes in the
point source catalog.

\begin{figure*}
\begin{center}
\leavevmode \epsfysize=8cm \epsfbox{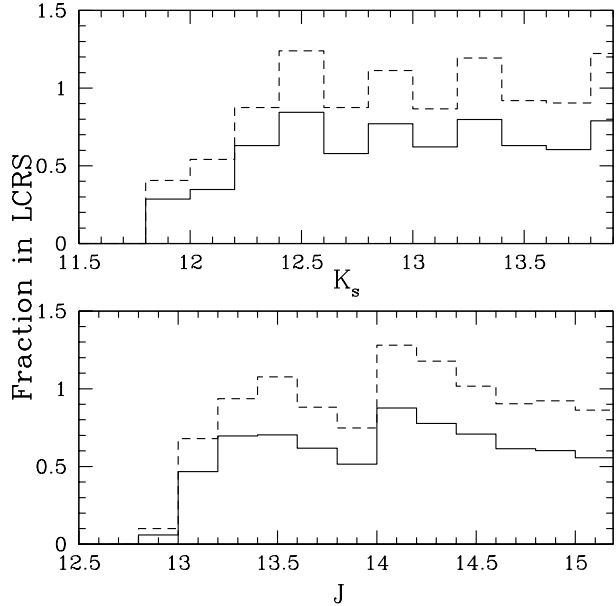}
\end{center}
\caption{The completeness of the LCRS, relative to 2MASS, as a function
of $J$  and $K_s$ magnitudes.  The solid line shows
the fraction of galaxies in the 2MASS catalog that have a corresponding
match in the LCRS spectroscopic catalog.  The dotted line shows the result
of weighting the LCRS counts to account for the sampling fraction.  The incompleteness
at the bright end is due to the bright magnitude cutoff of the LCRS.  
\label{fig-comp_lcrs}}
\end{figure*}

It is important to check the overall completeness
of the matched LCRS-2MASS catalog used in our analysis.
In Figure \ref{fig-comp_lcrs} we show the fraction of the galaxies in 2MASS for
which a successful match is made with a galaxy in the LCRS.  We
restrict our comparison to the largest ($20^\circ \times 1.5^\circ$)
contiguous patch covered by both surveys, which includes about 1300
galaxies.  The LCRS is not intended to be complete, but is a random
sample of a magnitude-limited survey.  The dashed line shows the
effect of applying the weights necessary to correct for this
incompleteness.  At $K_s>12.2$ and $J>13.2$, the corrected
completeness is close to 100\%.  At brighter magnitudes, the
completeness drops, due primarily to the LCRS's upper magnitude cutoff 
($R \sim 15$); we therefore limit the sample to magnitudes fainter than this
limit.  Cole \etal\ (\shortcite{Cole-2mass}) show that 2MASS is highly
complete and is not missing a substantial number of low surface brightness galaxies.
By association there can be 
no significant incompleteness in low surface brightness galaxies
in the matched LCRS-2MASS catalog, at least to
the relatively bright lower magnitude limit of 2MASS.  

The 2MASS catalog is complete to $K_s=13.2$, and $J=14.5$, as shown by
Cole \etal (\shortcite{Cole-2mass}).  Up to about 0.5 mag fainter than
these limits, the incompleteness is primarily due to misclassification
of faint galaxies as stars and small biases in the magnitude
measurements.  We mitigate some of this incompleteness by using
the LCRS to identify galaxies that 2MASS has classified as
stars.  Furthermore,
because our sample is considerably smaller than that of Cole \etal,
we are less concerned with systematic biasses of order $\lesssim0.1$ mag,
and we can maximize our useful sample size by
considering fainter magnitude limits.

\begin{figure*}
\begin{center}
\leavevmode \epsfysize=8cm \epsfbox{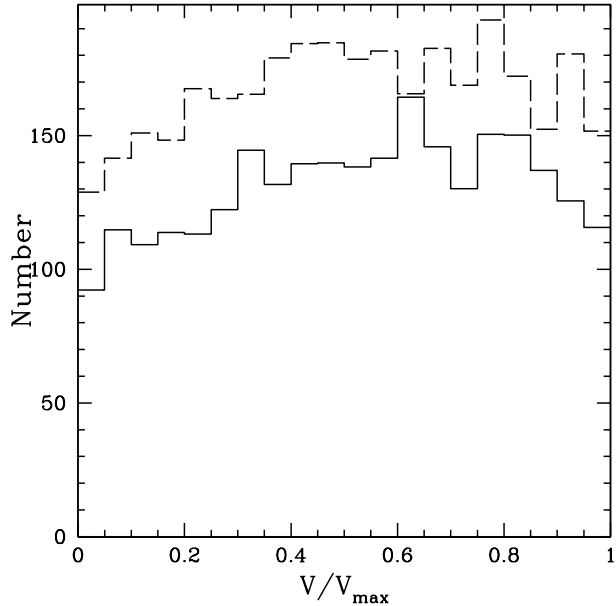}
\end{center}
\caption{The distribution of $V/V_{\rm max}$, for the $K_s$ band sample {\it (solid line)}, limited at $K_s=13.7$, and
the $J$ band sample {\it(dashed line)} limited at $J=15.0$.  The mean value is 0.521$\pm$0.006 for the $K_s$ sample and 0.513
$\pm$0.005 for the $J$ sample.
\label{fig-vvmax}}
\end{figure*}

One other way to test the completeness of our sample is
by considering the distribution of $V/V_{\rm max}$. For a galaxy at
redshift $z$, $V(z)$ is the volume between it and the lowest redshift $z_{\rm min}$ at which it would
still have been selected in the sample (due to the bright magnitude limit).  $V_{\rm max}$
is the volume between $z_{\rm min}$ and $z_{\rm max}$, where $z_{\rm max}$ is the redshift
at which the galaxy would drop out of the sample due to the faint magnitude limit.
For a uniform spatial distribution of 
objects the mean value of $V/V_{\rm max}$ is 0.5. Although galaxies 
are clustered, we expect that over the large volume surveyed here
the distribution can be thought of as uniform. In Figure \ref{fig-vvmax} we show the distribution
of $V/V_{\rm max}$ for the $K_s$ band sample, limited at $K_s=13.7$, and
the $J$ band sample, limited at $J=15.0$. 
With these limits, the distribution of $V/V_{\rm max}$ is approximately
uniform, though there is a systematic variation of about 15\%.
The mean value of $V/V_{\rm max}$ is about 3$\sigma$
larger than the value of 0.5,
which it would be for a statistically complete sampling of a spatially
uniform population.  This appears to be partly due to a deficit of galaxies with
$V/V_{\rm max}<0.3$, perhaps a consequence of Malmquist bias becoming important at $K_s>13.5$.
Indeed, moving the 
magnitude limit brighter by 0.2 mag does make the distribution of $V/V_{\rm max}$ more
uniform.  
None of the trends we discuss in this paper are affected by
decreasing the magnitude limit, although the uncertainties 
are increased and,
consequently, the significance of the results is diminished.
To maximize the useful sample size, we therefore restrict the $K_s$ sample to 
$12.2<K_s<13.7$, which consists of 2673 galaxies
(including 274 cluster galaxies and 954 group galaxies).  In $J$, we consider the sample of
3408 galaxies with $13.2<J<15.0$ (including 346 cluster galaxies and 1202 group galaxies).

\subsection{Potential Biases and Systematic Uncertainties}\label{sec-bias}
Clusters and groups (associations) are defined by identifying friends-of-friends in redshift
space. Only associations with three or more members are retained
in the final catalog.   A concern is that near this cutoff, 
associations with steeper faint end slopes $\alpha$, or brighter characteristic magnitudes, might
be preferentially included in the catalog because more member galaxies would
be above the magnitude limit. This could bias the final LF;
however, we expect the overall effect to be small because the majority of the associations
in our catalogue have more than the minimum three members, and are unlikely
to drop below this threshold given reasonable variations in the LF.
To test for this potential effect, we have constructed mock
catalogs, as described in Christlein (\shortcite{Christ}), by adding
$\sim$350,000 artificial associations to the LCRS catalog.  In one
catalog, the associations have LFs with steep faint end slopes of
$\alpha=-1.2$, and in another they have shallow slopes, $\alpha=-0.6$.
As expected, we find  no evidence from this experiment that the recovery fraction
of associations with steeper faint-end slopes is higher than for those with
flatter slopes.  Thus, this potential bias does not have
a significant effect on our selection, even when considering mock samples
containing many more galaxies than the actual data.

\begin{figure*}
\begin{center}
\leavevmode \epsfysize=8cm \epsfbox{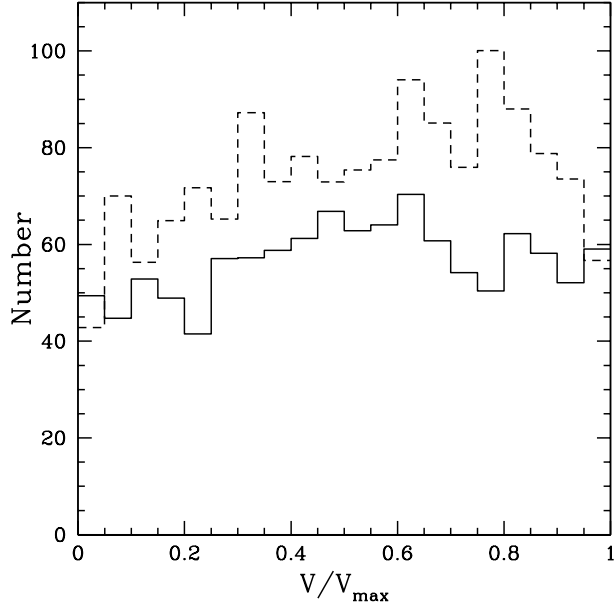}
\end{center}
\caption{For the $K_s$ selected sample, the distribution of $V/V_{\rm max}$ for emission line galaxies {\it (solid line)} and
NEL galaxies {\it(dashed line)}.  The mean value is 0.515$\pm$0.008 for the emission line galaxies
and 0.525$\pm$0.008 for the NEL galaxies.
\label{fig-vvmax2}}
\end{figure*}

There is a potential bias in any redshift catalog to be more complete
in emission line galaxies because it is easier to measure redshifts
for these galaxies than for non-emission line galaxies.  This
bias is particularly acute for the faintest galaxies in the
sample.  The magnitude ranges of the LCRS and 2MASS are not well
matched, so only the brightest galaxies in the LCRS are included in
our sample.  In particular, most of the galaxies have $15<R_c<16.5$,
well above the completeness limit of the LCRS.  Because the magnitude
range is small, and because we are far from the magnitude limit of the
LCRS, there is little danger that a weak magnitude-dependent
completeness function of the NEL galaxies has much effect on our
analysis.  In Figure \ref{fig-vvmax2} we show the $V/V_{\rm max}$
distributions for the NEL and EL galaxies in the $K_s$
selected sample.  For the emission line galaxies, $\langle V/V_{\rm
max}\rangle=0.515\pm0.008$, while for the NEL galaxies $\langle
V/V_{\rm max}\rangle=0.525\pm0.008$.  There is no evidence that
the completeness of the EL sample differs from that of the NEL sample.

\section{Results}\label{sec-lfunc}
We compute the luminosity functions using both the
non-parametric method of Efstathiou, Ellis \& Peterson (\shortcite{EEP}, EEP) 
and by assuming a Schechter (\shortcite{S76}) function shape
as described by Sandage, Tamman \& Yahil (\shortcite{STY}, STY).  
Luminosities are calculated assuming an $\Omega=1$, $\Lambda=0$ cosmology, with the
Hubble constant parametrized as $H_\circ=100 h$ km s$^{-1}$ Mpc$^{-1}$.  Where not
explicitly shown, we use $h=1$.
K-corrections are taken from
Poggianti (\shortcite{P97}); they are small (less than 0.1 mag at $z=0.1$) 
and nearly independent of galaxy type.  We neglect evolutionary corrections
and extinction effects, both of which affect the magnitudes at less than
the 0.1 mag level (e.g., Cole \etal\ 2000b).

We do not compute volume densities to normalize the luminosity functions.
An estimate of the global
normalization has already been presented in Cole \etal (\shortcite{Cole-2mass}).  For the field galaxies and the
total sample, we normalize the luminosity functions to the weighted number of galaxies brighter than
$M_{K_s}=-21.5$ and $M_J=-20.5$.  For the group and cluster 
environments, we divide by the number of associations to obtain
the number of galaxies per group or cluster that are brighter than this limit.  

\begin{figure*} \begin{center} 
\leavevmode \epsfysize=8cm \epsfbox{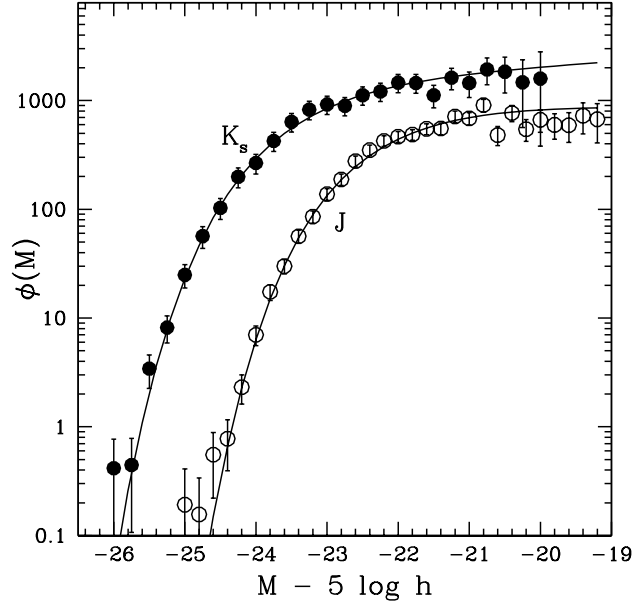} 
\end{center} \caption{The $J$ {\it (open symbols)}
and $K_s$-band {\it (filled symbols)} luminosity functions.  The
points and solid lines are the determinations from the EEP and STY
methods, respectively.  The luminosity functions are arbitrarily
normalized to the weighted number of galaxies brighter than
$M_{K_s}=-21.5$ and $M_J=-20.5$ (the $J$-band LF is arbitrarily
shifted vertically for clarity).  \label{fig-lfunc}} \end{figure*}

\begin{figure*} \begin{center} 
\leavevmode \epsfysize=8cm \epsfbox{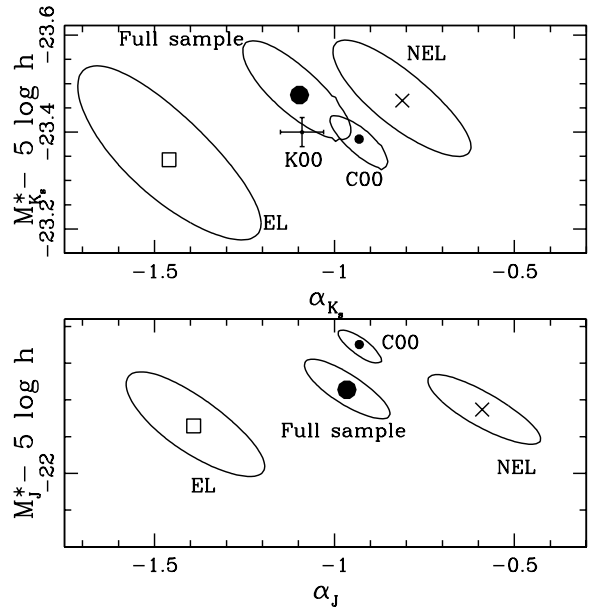} 
\end{center} \caption{The error ellipses (67\%
confidence contours) on the Schechter parameters of the luminosity
functions, for the full sample {\it(filled circle)}, EL galaxies {\it(square)} and
NEL galaxies {\it(cross)}.
Recent determinations by Kochanek et al. (2000) and Cole et al. (2000b)
are marked K00 and C00 respectively.  The small difference between the
C00 and K00 results are due to the different magnitude systems used.
All parameters are for $\Omega=1$, $\Lambda=0$, and include only
$k-$corrections, with no evolution.  \label{fig-errellipse}}
\end{figure*}

In Figure \ref{fig-lfunc} we show the $J$ and $K_s$-band luminosity
functions for the full sample.  Both are well fit by Schechter
functions with parameters \mstar$=-23.48+5\mbox{log} h$,
\kalpha$=-1.10$, and \jmstar$=-22.23+5\mbox{log} h$, \jalpha$=-0.96$.
In the non-parametric method, the uncertainties are determined from
the information matrix (EEP\nocite{EEP}).
The uncertainties in the Schechter parameters are
determined from contours of the likelihood function,
and are shown as 67\% confidence elliptical contours in Figure
\ref{fig-errellipse}.  In Figure \ref{fig-errellipse}, we compare
our result with recent measurements by Kochanek \etal
(\shortcite{Kochanek-KLF}) and Cole \etal (\shortcite{Cole-2mass}).
As shown in Cole \etal, the 
small difference between their result and that of Kochanek \etal\ 
is mostly due to the difference in the magnitude
definition (we use that of Cole \etal).  For the $K_s$ sample, we are
in good agreement with Cole \etal\  For the $J$ sample, 
the 1$\sigma$ contours do not overlap, which 
may reflect the incompleteness introduced by including galaxies
0.5 magnitudes fainter than the nominal 2MASS completeness limit.  We recover
the Cole \etal\ result if we impose the same magnitude limits.  While
statistically significant, the difference in the $J$ band results is
small, and does not affect our conclusions based on subdividing the sample
according to spectral type and environment.

\subsection{Dependence on Spectral Type}\label{sec-sptype} 
\begin{figure*} \begin{center} 
\leavevmode \epsfysize=8cm \epsfbox{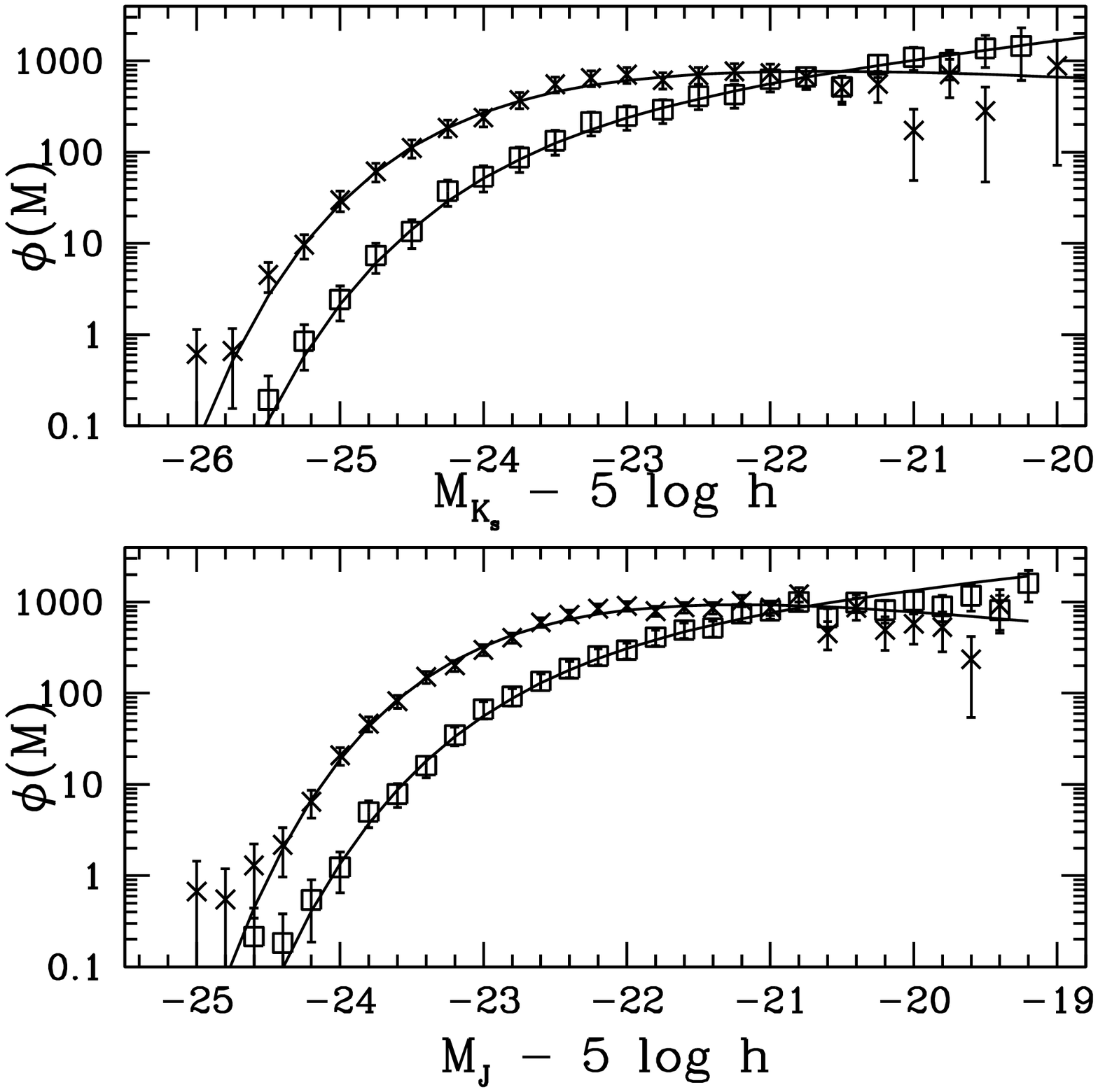} 
\end{center} \caption{The $J$ and $K_s$-band
luminosity functions divided by spectral type.  Open squares represent
emission line galaxies, while crosses represent non-emission line
galaxies.  The luminosity functions are arbitrarily normalized to the
weighted number of galaxies brighter than $M_{K_s}=-21.5$ and
$M_J=-20.5$.  \label{fig-elines}} \end{figure*}

In Figure
\ref{fig-elines} we present the LFs divided by spectral
type.  In $J$ and $K_s$ the NEL galaxies have a significantly flatter
faint-end slope than the EL galaxies; this result is also
evident from the error ellipses in Figure \ref{fig-errellipse}.

\begin{figure*} \begin{center} 
\leavevmode \epsfysize=8cm \epsfbox{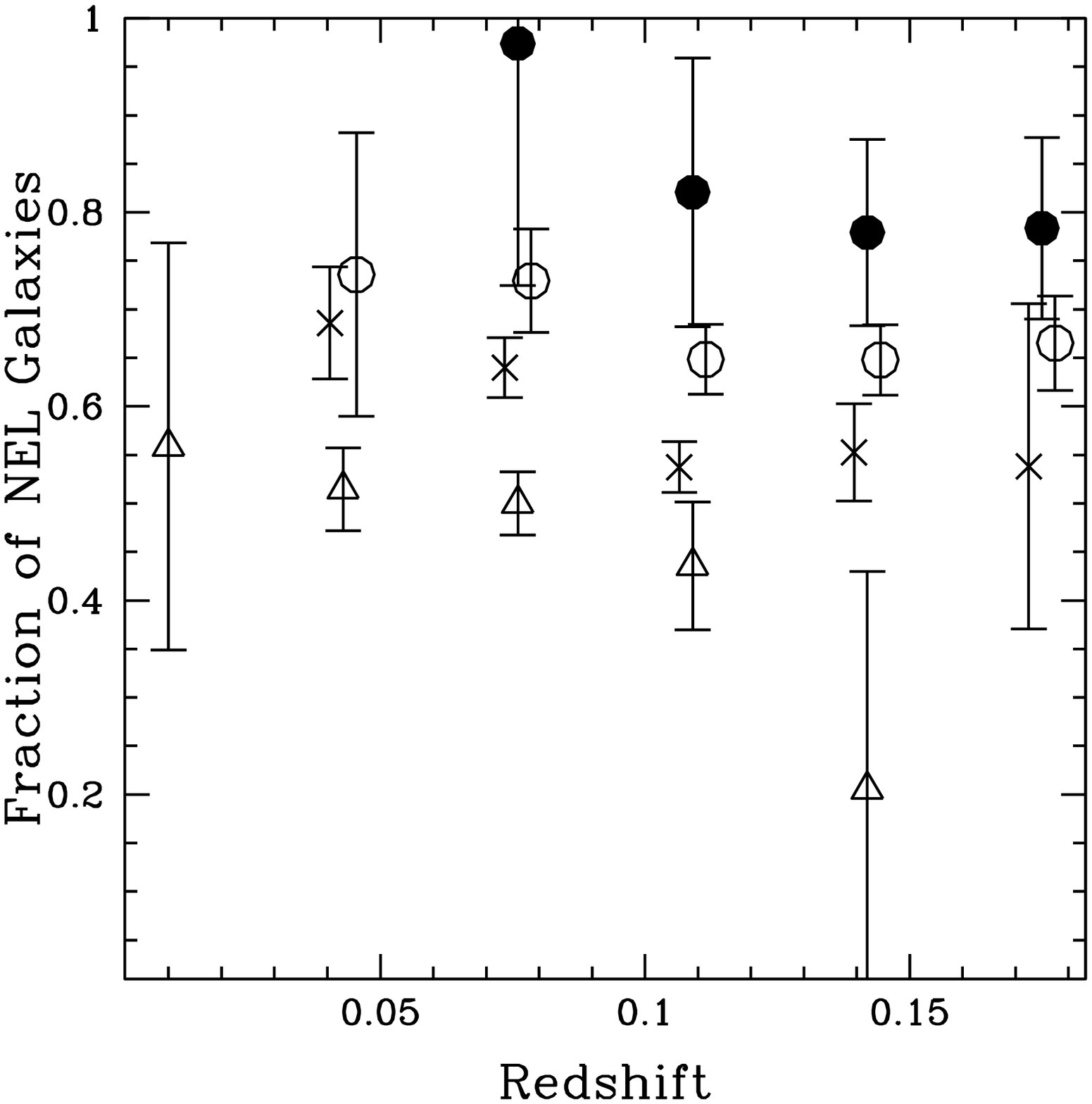} 
\end{center} \caption{The fraction of NEL
galaxies as a function of redshift is shown in the for four luminosity
bins: $-25.5<M_{K_s}<-24.5$ (filled circles); $-24.5<M_{K_s}<-23.5$
(open circles); $-23.5<M_{K_s}<-22.5$ (crosses); and
$-22.5<M_{K_s}<-21.5$ (triangles).  \label{fig-apeffect}}
\end{figure*}

Kochanek (\shortcite{Kochanek-LCRS}) has recently suggested that the
fixed angular size of the fibers used to obtain the spectra
leads to a substantial aperture related bias.  To address the significance
of this effect we show the fraction of
NEL galaxies as a function of redshift, $f_{\rm NEL}(z)$, in four bins
of absolute magnitude (Figure \ref{fig-apeffect}).
There is a weak trend, in the expected sense
that more distant galaxies are more likely to have spectra with
emission lines because the fixed aperture includes more of the galaxy
(we have already shown, in \S\ref{sec-bias}, that this is not due to 
a redshift dependent incompleteness in NEL galaxies).
Although low luminosity galaxies tend to be physically smaller than
high luminosity galaxies, they also lie at preferentially lower redshifts
in a magnitude-limited sample.  Thus, we expect the two effects to
partially cancel out and reduce the size of this bias.

\subsection{Environmental Dependence}\label{sec-env} 
\begin{figure*} \begin{center} 
\leavevmode \epsfysize=8cm \epsfbox{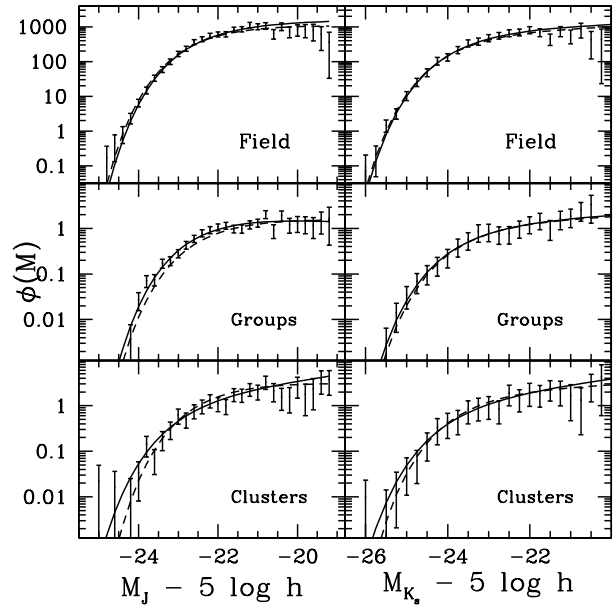} 
\end{center} \caption{The $J$ and $K_s$-band
luminosity functions as a function of environment, shown as the error
bars (EEP method) and the {\it solid lines} (STY method).  
The luminosity functions are arbitrarily
normalized to the weighted number of galaxies brighter than
$M_{K_s}=-21.5$ and $M_J=-20.5$, divided by the number of groups or
clusters as appropriate in the lower two panels.  The {\it dashed
lines} show the (arbitrarily renormalized) luminosity function for the
whole sample for comparison.  In most cases, this curve is hidden by
the overlapping luminosity functions.  \label{fig-env_lf}}
\end{figure*}

\begin{figure*} \begin{center} 
\leavevmode \epsfysize=8cm \epsfbox{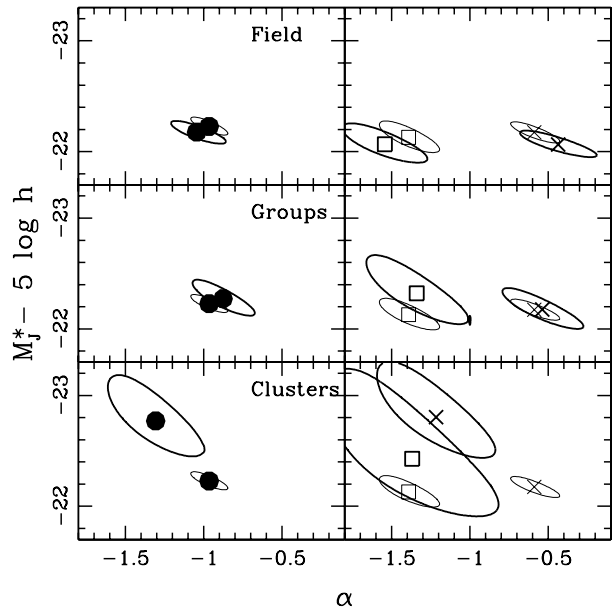} 
\end{center} \caption{The 67\% confidence
error ellipses of the $J$-band Schechter fits for galaxies divided by
environment.  On the {\it left} are results for galaxies in the field
({\it top}), group ({\it middle}) and cluster ({\it bottom}) samples.
The heavy ellipses correspond to the Schechter parameters for that
bin; the lighter ellipses correspond to the Schechter parameters for
the total sample, reproduced in each panel for comparison.  On the
{\it right} side of the figure, the samples are divided into EL
and NEL galaxies.  The best fit values for
the emission line galaxies are indicated with a square, and the values
for the NEL galaxies with a cross.  All parameters are for $\Omega=1$, $\Lambda=0$, and include only
$k-$corrections, with no evolution.  \label{fig-jenv_sty}}
\end{figure*}

\begin{figure*} \begin{center} 
\leavevmode \epsfysize=8cm \epsfbox{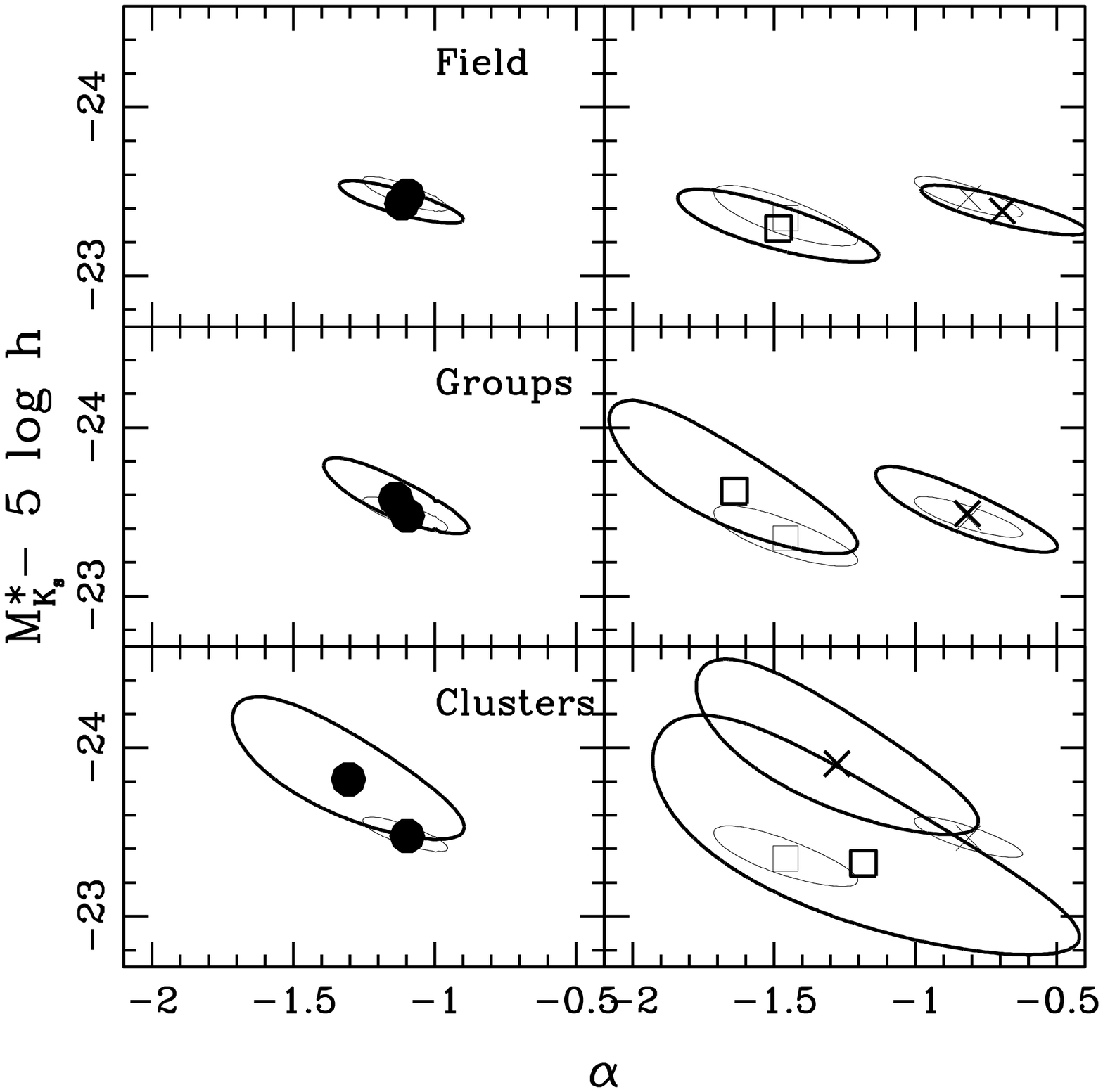} 
\end{center} \caption{As Figure
\ref{fig-jenv_sty}, for the $K_s$ band data.  \label{fig-env_sty}}
\end{figure*}

As described in
\S\ref{sec-data}, we divide our sample into three environmental
categories, corresponding to field, groups, and clusters.  In
Figure \ref{fig-env_lf} we show the luminosity functions of these
three classes of galaxies, compared with the best fit Schechter
functions for the full sample; the 67\% confidence error ellipses  on the Schechter
parameters are shown in Figures \ref{fig-jenv_sty} and
\ref{fig-env_sty}.  The 
Schechter parameters
of the $J-$band cluster LF are inconsistent with those of the field LF at
the 95\% confidence level, as determined from the likelihood contours.
The trend is more significant in the $J-$band due to the larger sample size
as a consequence of the deeper photometry.
The difference is in the sense that the cluster LF has a brighter \jmstar\
and a steeper $\alpha_J$; however, given the degeneracy between these parameters
it is not clear precisely how the LF shape varies. 
If we restrict the Schechter fit to a limited luminosity range (e.g., 
brighter than $M_J=-20.5$, or fainter than $M_J=-23.5$), the
best-fit parameters change along the major axis of the error ellipse.  However,
both the cluster and field parameters move in the same direction, and the difference
between them is approximately maintained.  It appears, therefore, that 
the environmental dependence of the LF is not restricted to just the bright or
faint end.


The differences among the LFs of different environments are larger
when only the NEL galaxies are considered, as shown in Figure
\ref{fig-jenv_sty}.  In the field, the LF of NEL galaxies is much
flatter than that of the full sample, but in
clusters the LFs of EL and NEL are statistically indistinguishable. 
We discuss this
result further in \S\ref{sec-spec}.  
The Schechter parameters of the $J$ and $K_s$ luminosity functions,
for each data subsample, are tabulated in Table \ref{tab-params}.

\begin{table*} \begin{center} \begin{tabular}{llcc|cc} \hline
Environment&Spectral Type &\mstar&\kalpha&\jmstar&\jalpha\\ \hline
Total   &all& -23.48$\pm$0.08 & -1.10$\pm$0.14 & -22.23$\pm$0.07 & -0.96$\pm$0.12 \\
        &EL & -23.34$\pm$0.16 & -1.46$\pm$0.26 & -22.13$\pm$0.14 & -1.39$\pm$0.19 \\
        &NEL& -23.46$\pm$0.12 & -0.81$\pm$0.28 & -22.17$\pm$0.09 & -0.59$\pm$0.15 \\ \hline
Field   &all& -23.43$\pm$0.12 & -1.12$\pm$0.21 & -22.18$\pm$0.10 & -1.04$\pm$0.18 \\
        &EL & -23.28$\pm$0.20 & -1.48$\pm$0.36 & -22.07$\pm$0.17 & -1.54$\pm$0.27 \\
        &NEL& -23.38$\pm$0.16 & -0.69$\pm$0.28 & -22.06$\pm$0.10 & -0.44$\pm$0.24 \\ \hline
Groups  &all& -23.58$\pm$0.13 & -1.14$\pm$0.26 & -22.37$\pm$0.17 & -0.88$\pm$0.20 \\
        &EL & -23.62$\pm$0.44 & -1.64$\pm$0.43 & -22.32$\pm$0.29 & -1.34$\pm$0.34 \\
        &NEL& -23.49$\pm$0.26 & -0.81$\pm$0.33 & -22.18$\pm$0.17 & -0.54$\pm$0.27 \\ \hline
Clusters&all& -23.81$\pm$0.40 & -1.30$\pm$0.43 & -22.77$\pm$0.34 & -1.30$\pm$0.32 \\
        &EL & -23.31$\pm$0.70 & -1.18$\pm$0.76 & -22.43$\pm$0.65 & -1.37$\pm$0.56 \\
        &NEL& -23.91$\pm$0.52 & -1.28$\pm$0.50 & -22.80$\pm$0.43 & -1.22$\pm$0.39 \\ \hline \end{tabular}
\end{center} 
\caption{We use the cosmological parameters 
$\Omega=1$, $\Lambda=0$, $h=1$.  The tabulated 1$\sigma$ errors are strongly correlated, as shown by the 
error ellipses in Figures
\ref{fig-errellipse}, \ref{fig-jenv_sty} and \ref{fig-env_sty}.  \label{tab-params}} 
\end{table*}
\section{Discussion}\label{sec-discussion} \subsection{Comparison with
Literature}\label{sec-lit} It is encouraging that the global
luminosity functions measured here, based on redshifts from the LCRS,
are in such good agreement with recent determinations by Cole \etal\
(\shortcite{Cole-2mass}) and Kochanek \etal\
(\shortcite{Kochanek-KLF}). As shown by those authors, all infrared LFs 
derived from large samples are consistent
(e.g., \cite{MSE,GPMC,Gardner,SSCM,Loveday-klf}).  This concordance is
in contrast with the historical situation at optical wavelengths, where there has
been considerable disagreement between results in the literature.  In
particular, optical LFs derived from the LCRS
(\cite{L+96,Christ}) are generally shallower than most other
measurements, 
and it has recently been shown that this difference
results from an underestimation of galaxy magnitudes due to the
shallow isophotal limit of the LCRS photometry and from the 
exclusion of low central surface brightness galaxies
(Blanton \etal \shortcite{Sloan_lf}).  
The fact that we do not find a shallow
slope in the infrared LF supports the interpretation that,
at least over the 
relatively narrow, bright magnitude range of 2MASS,
it is the $R$ magnitudes, and not the survey selection itself, that is responsible
for the discrepancy in the optical.

Analysis of the optical LF dependence on environment was done by
Christlein (\shortcite{Christ}), using the LCRS sample.  Although the
LCRS magnitudes are systematically biased due to the bright isophotal
limit, this bias applies uniformly to the whole sample and is therefore
unlikely to affect the trends discussed in that paper.  Christlein
found strong evolution of the LF with environment, with the Schechter
parameters changing along the major axis of the ellipse, toward
steeper $\alpha$ and brighter $M^\ast$ with larger $\sigma$.  The
magnitude and sense of this change are fully consistent with the
results presented in \S\ref{sec-env}.  The fact that Christlein finds
such a continuous dependence on $\sigma$ (something we are unable to
investigate due to our smaller sample size) provides encouraging
support that the effect is real.  Zabludoff \& Mulchaey
(\shortcite{ZM00}) found similar evidence for steep, bright $R$-band
LFs in X-ray selected groups ($\alpha=-1.3$). Because these groups generally have
$kT\gtrsim 1$ keV, many would be considered ``clusters'' by our criteria,
and their results for steeper optical LFs in these systems are very
similar to our results for the cluster infrared LFs. 

Previously published evidence for an environmental dependence of the
infrared LF is not as strong as for the optical LF.  Both De Propris
\etal\ (\shortcite{Roberto98}) and Andreon \& Pell{\'o}
(\shortcite{AP00}) find a steep faint end slope in the $H$ band LF of
the Coma cluster, but only at magnitudes fainter than our sample
probes.  Trentham \& Mobasher (\shortcite{TMo98}) find no evidence for
a difference between the $K$-band LFs of cluster and field galaxies,
over $-24<M_K<-22$, although the uncertainties are large.  Our sample
is the first that is large enough to detect this weak effect at these
luminosities.
We look forward to dramatically
improving the precision of this result when the full 2MASS catalog is released.

\subsection{The Stellar Mass Function} 
\begin{figure*} \begin{center} 
\leavevmode \epsfysize=8cm \epsfbox{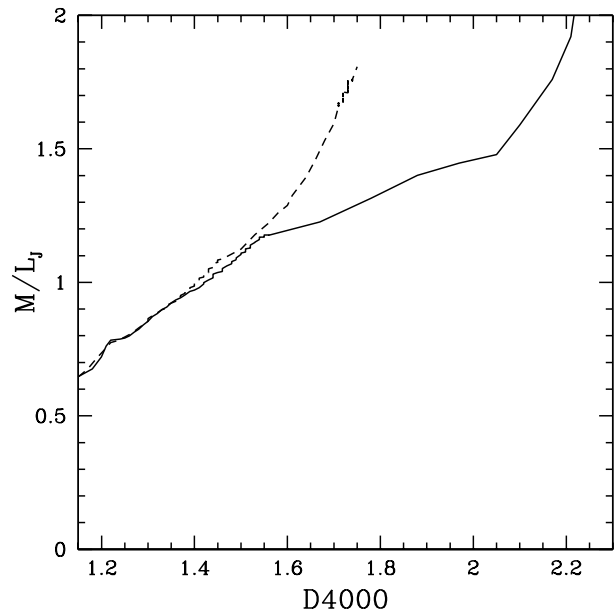} 
\end{center} \caption{Mass-to-light ratio in the $J$
band, as a function of $D_{4000}$, for a Kennicutt (1983) initial mass function from the
models of Fioc \& Rocca-Volmerange (1997).  The solid line is a single
burst population, and the dashed line represents a galaxy with
constant star formation rate.  In both models, the galaxy evolves from
left to right with time, reaching its maximum $D_{4000}$ at 13 Gyr.
\label{fig-MLR}} \end{figure*}

Do the observed dependencies of the
infrared LF on environment and spectral type necessarily imply an
analogous difference in the stellar mass functions?  The infrared light is a
good, but not perfect tracer of the stellar mass; the mass-to-light
ratio (M/L) can be expected to increase by about a factor of two
between early- and late-type galaxies (e.g., \cite{BdJ_ML}).  
To estimate M/L, constraints on the mean stellar age are
required.  The optical-IR colors of our sample are unreliable due to
the shallow isophotal limit of the LCRS, and the IR colors alone are
insensitive to stellar population differences. The best indicator
available to us is therefore the $D_{4000}$ spectral index from the
LCRS spectra.  
We use the solar-metallicity models
of Fioc \& Rocca-Volmerange (\shortcite{PEGASE}) to generate spectra
from a given star formation history.  We then compute $D_{4000}$ directly
from the spectra, using the same definition as Zabludoff \etal
(\shortcite{Z+96}). The correlation between $M/L_J$ and $D_{4000}$ depends
only weakly on star formation history. Figure \ref{fig-MLR} shows
model results for a single burst, simple stellar
population (SSP) and for a galaxy with a constant star formation
rate, using a Kennicutt (\shortcite{K83}) initial mass function.  
The two models can differ by as much as 40\%, over a narrow range of $D_{4000}$,
but the ranking with $M/L$ is unchanged.  
$D_{4000}$ is sensitive to metallicity and reddening, which we neglect,
but is less sensitive than optical colors.
For example, an extinction of $A_{\rm V}=1.5$ magnitudes only
reddens $D_{4000}$ by about 0.2.

\begin{figure*}
\begin{center}
\leavevmode \epsfysize=8cm \epsfbox{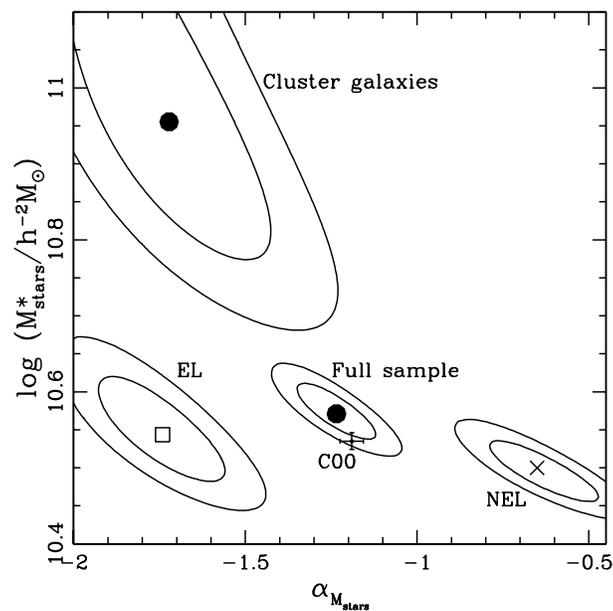}
\end{center}
\caption{67\% and 95\% confidence ellipses on the Schechter parameters for
various stellar mass functions, using $J$-band mass-to-light ratios determined
from the $D_{4000}$ index, as described in the text.  The point with error
bars is the result of Cole \etal\ (2000b) using the Kennicutt (1983)
initial mass function.  The differences in luminosity function shape when divided by
spectral type and environment are even more pronounced in the stellar mass
functions than in the luminosity functions.  \label{fig-mfunc}}
\end{figure*}

In Figure \ref{fig-mfunc} we show the 67\% and 95\% error ellipses on the Schechter
parameters for the stellar mass functions derived using the M/L
corrections derived from the SSP model.  The differences observed in the
luminosity functions are {\it exacerbated} in the stellar mass functions.
Thus, we conclude that the stellar mass function is not
universal, but depends on both galaxy type and the local environment.

\subsection{Implications and Speculations}\label{sec-spec} Both
Christlein (\shortcite{Christ}) and Zabludoff \& Mulchaey
(\shortcite{ZM00}) find that the change in the optical LF shape with
environment is due almost exclusively to the NEL galaxies.  We confirm
that the difference in the infrared cluster LF, relative to the field,
is dominated by this population.  
It is interesting that
the faint end slope of the NEL galaxies in clusters, as parametrized
by $\alpha$, is similar to that of the overall field faint end
slope.  
One interpretation for this  similarity is
that the bulk of the cluster population is built up 
by accreting field galaxies with little effect other than
the cessation of star formation, as modeled for example
by Balogh, Navarro \& Morris (\shortcite{infall}). However, from only the agreement
of the LFs, we cannot exclude scenarios in which the similar
initial progentors of faint galaxies in various environments
began forming stars 
at different times (the cluster ones earlier by virtue of
originating near a large mass perturbation). In the latter scenario,
one expects the field LF to evolve to the cluster LF in time,
as gas is consumed by galaxies and their star formation stops. 

\begin{figure*}
\begin{center}
\leavevmode \epsfysize=8cm \epsfbox{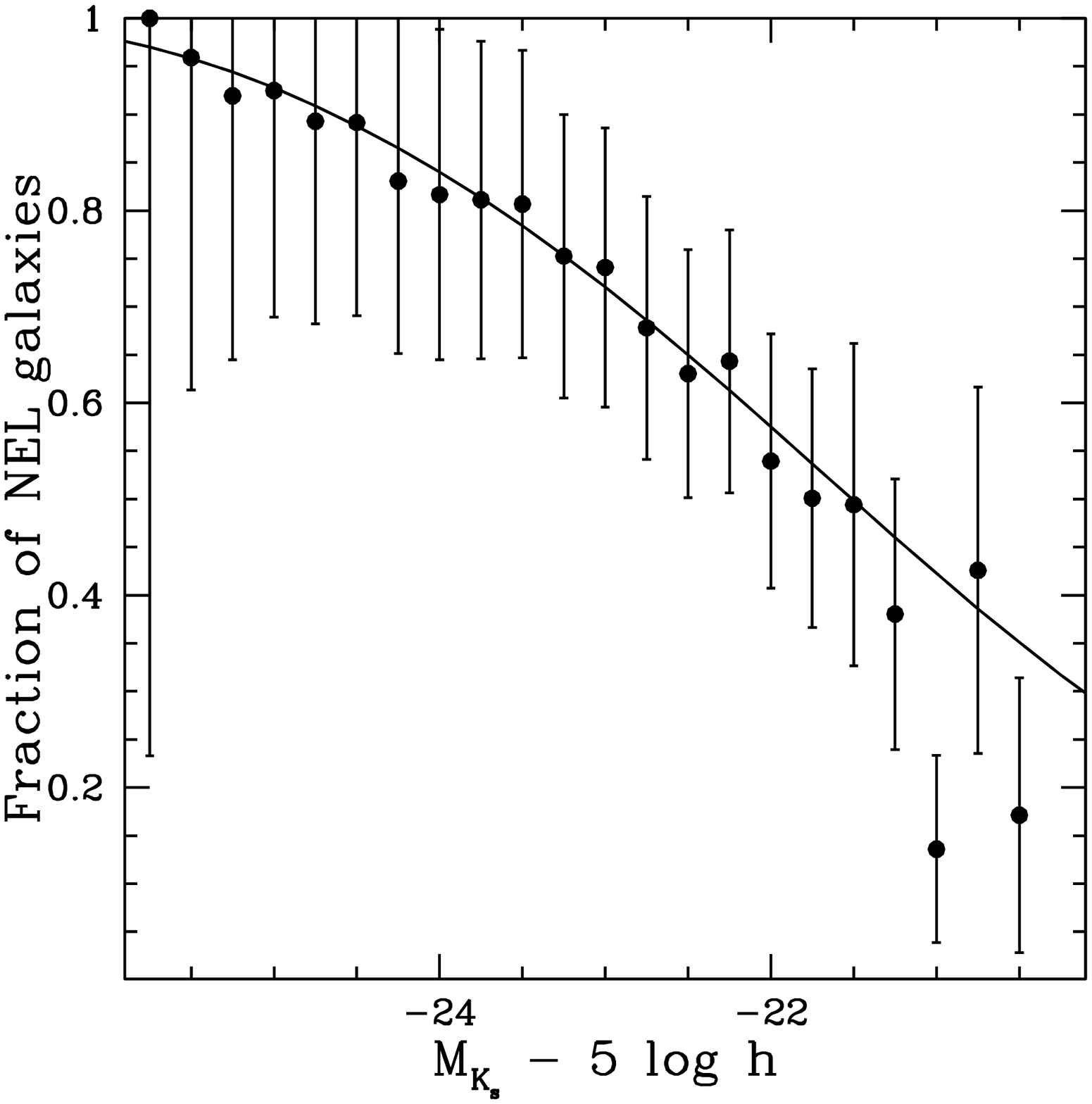}
\end{center}
\caption{The fraction of non-emission line galaxies as a function of luminosity, as determined
by the non-parametric luminosity function ({\it points}) and the Schechter function fits
({\it solid line}).
\label{fig-fnel}}
\end{figure*}

In Figure \ref{fig-fnel} we show the fraction of NEL
galaxies as a function of luminosity for the global sample.  This
Figure shows a strong trend that, unfortunately, is difficult to
interpret because metallicity and extinction also
correlate with luminosity in a way that makes [OII] stronger in low
luminosity galaxies (e.g., Jansen \shortcite{Jansen01}). 
We cannot infer the extent to which this relation indicates an inherent
correlation between galaxy mass and instantaneous star formation rate ---
a correlation that  would be an important test of galaxy formation models.  In
particular, a variation in the fraction of currently star forming
galaxies with mass could indicate that the dominant mode of star formation
(i.e., burst-like or continuous) for a galaxy in the field is mass dependent (e.g., Kauffmann,
Charlot \& Balogh, in preparation).   

\section{Conclusions}\label{sec-conc}
We have measured the $J$ and $K_s$-band galaxy luminosity functions from the
2MASS, using redshifts from the LCRS, and divided the sample based on
spectral type and environment.  We draw the following conclusions:
\begin{itemize} 
\item In non-cluster environments,
the infrared luminosity function of emission
line (EL) galaxies is steeper than that of non-emission line (NEL)
galaxies.
\item There is a statistically 
significant difference between the
cluster and field galaxy infrared luminosity functions, in the
sense that $M^\ast$ is brighter, and the faint end slope $\alpha$ is steeper, in
clusters.
\item Using the $D_{4000}$ spectral index from the 
LCRS spectra to aid us in
converting from infrared luminosity to stellar mass, we conclude that
the results just described also hold for the stellar mass functions and,
in fact, the differences are more pronounced.
\item The shape of the NEL galaxy stellar mass
function in clusters is similar to that of the global
stellar mass function in the field.  One explanation of this trend is that the cluster
is mostly built up of field EL galaxies, in which star formation has ceased
either because (1) the galaxies consumed all their gas (either through accelerated
star formation due to the cluster environment, or by forming earlier due to their
proximity to a large mass perturbation); or
(2) they lost some or all their gas through processes like ram pressure stripping.

\end{itemize}

\section*{Acknowledgements}
We thank Guinevere Kauffmann for suggesting the idea of comparing the
2MASS and LCRS catalogs.  We are also grateful to the LCRS collaboration
for making the spectra available.
MLB acknowledges support from a PPARC
rolling grant for extragalactic astronomy and cosmology at Durham.
DC and AIZ acknowledge financial support from NASA grant GO0-1007X.
DZ acknowledges financial support from a NSF CAREER grant (AST-9733111),
and a fellowship from the David and Lucile Packard Foundation.

\bibliographystyle{astron_mlb}
\bibliography{ms}

\end{document}